\documentclass[12pt]{article}
\usepackage[english]{babel}
\usepackage[T1]{fontenc}
\usepackage{amsfonts}
\begin{document}
\selectlanguage{english}

\baselineskip 0.76cm
\topmargin -0.4in
\oddsidemargin -0.1in

\let\ni=\noindent

\renewcommand{\thefootnote}{\fnsymbol{footnote}}

\newcommand{\SM}{Standard Model }

\pagestyle {plain}

\setcounter{page}{1}



~~~~~~
\pagestyle{empty}

\begin{flushright}
IFT-- 08/13
\end{flushright}

\vspace{0.6cm}

{\large\centerline{\bf Hidden-sector fermions in two or three generations.}}

\vspace{0.5cm}

{\centerline {\sc Wojciech Kr\'{o}likowski}}

\vspace{0.23cm}

{\centerline {\it Institute of Theoretical Physics, University of Warsaw }}

{\centerline {\it Ho\.{z}a 69,~~PL--00--681 Warszawa, ~Poland}}

\vspace{0.6cm}

{\centerline{\bf Abstract}}

\vspace{0.2cm}

We recall the argument based both on Dirac's square root procedure and an intrinsic 
Pauli principle that sterile fundamental fermions with spin 1/2 (sterinos) ought to appear 
in Nature in two or three generations, while the Standard Model leptons and quarks are forced to 
develop three generations. Then, sterinos in two or three generations, if stable, lead to the cold 
dark matter in two or three species, when the thermal freeze-out mechanism works properly. If 
they are stable only in their lowest generation, the so called exciting cold dark matter may be 
formed. 

\vspace{0.3cm}

\ni PACS numbers: 14.80.-j , 04.50+h , 95.35.+d 

\vspace{0.6cm}

\ni November 2008

\vfill\eject

~~~~~

\pagestyle {plain}

\setcounter{page}{1}

\vspace{0.3cm}

\ni {\bf 1. Introduction}

\vspace{0.3cm}

The global interpretation of recent, still preliminary results of direct ({\it e.g.} DAMA) and indirect ({\it e.g.} PAMELA and, perhaps, INTEGRAL) detection experiments for the cold dark matter [1] suggest that there may be more than one of its species with different masses [2]. Moreover, the higher-mass species may be not necessarily stable: they may be collisionally excited and decay afterwards, leading to the so called exciting cold dark matter [3].

In the first Ref. [4], where we considered sterile spin-1/2 fermions (sterinos) as the stuff whose thermal freeze-out relic provides today's cold dark matter, we pointed out that they may appear in more than one generation. This could be consistent with the recent interpretative suggestion about the multiple structure of cold dark matter.

At any rate, we can readily extend our detailed model of the hidden sector of the Universe described in Refs. [4], assuming that among its components, namely sterile {spin-$\!1\!/2$} fermions (sterinos) and spin-0 bosons (sterons), the former appear in two or more generations (then, in our formalism there are two or more dimensionless constants $\zeta$ as well as $\xi$, but still one dimensionless constant $\eta$ --- {\it cf.} the last Ref. [4] for the option with one sterino generation, and footnote $^*$ for the extension). The nonzero vacuum expectation value of steron field generates spontaneously sterino masses (as well as a steron mass). Interactions within the hidden sector are mediated by the massive antisymmetric-tensor field $A_{\mu\,\nu}$ (of dimension one), whereas this sector communicates with the \SM sector (after the electroweak symmetry is spontaneously broken by the \SM Higgs bosons) through the {\it photonic portal}. This portal is provided by the electromagnetic field $F_{\mu\,\nu} = \partial_\mu A_\nu - \partial_\nu A_\mu $ included (besides the sterino and steron fields, $\psi$ and $\varphi$) into the source of mediating field $A_{\mu\,\nu}$. A mass of quanta of the field $A_{\mu\,\nu}$ --- expected to introduce into our model a large mass scale --- is also generated spontaneously by the nonzero expectation value of the steron field. This nonzero value produces spontaneously sterino magnetic moments as well [4], which can make of the cold dark matter (composed of sterinos) a weakly magnetized medium in external magnetic fields (appearing locally in the Universe){\footnote{Our formalism for two or more sterino generations goes as follows. Before the spontaneous mass generation for sterinos, sterons and quanta of the field $A_{\mu \nu}$, the electromagnetic Lagrangian $-(1/4) F_{\mu\,\nu} F^{\mu\,\nu}  - j_\mu A^\mu $ supplemented by the hidden-sector Lagrangian with two or more sterino generations gets the form

\vspace{-0.3cm}

\begin{eqnarray*}
\cal{L}\!\!\!& = &\!\!\!-\frac{1}{4} F_{\mu\,\nu} F^{\mu\,\nu}  - j_\mu A^\mu -\frac{1}{2} \sqrt{f}\left(\varphi F_{\mu \nu} + \sum_i \zeta_i \bar{\psi}_i \sigma_{\mu \nu} \psi_i\right) A^{\mu\,\nu} \\ 
& &-\frac{1}{4}  \left[(\partial_\lambda A_{\mu\,\nu}) (\partial^\lambda A^{\mu\,\nu}) - \frac{1}{2}\eta\, \varphi^2 A_{\mu\,\nu} A^{\mu\,\nu}\right]  \\ 
& & + \sum_i\,\bar\psi_i \left(i \gamma^\lambda \partial_\lambda -  \xi_i  \varphi\right) \psi_i  + \left[\frac{1}{2}(\partial_\lambda\varphi) (\partial^\lambda\varphi ) - V(\varphi) \right] \;, 
\end{eqnarray*}

\vspace{-0.1cm}

\ni where $V(\varphi) \equiv -(1/2) \mu^2 \varphi^2 + (1/4) \lambda \varphi^4 $ and $\varphi \equiv <\!\!\varphi\!\!>_{\rm vac} + \varphi^{(\rm ph)}$. After the spontaneous mass generation by $<\!\!\varphi\!\!>_{\rm vac}$ the potential $V(\varphi)$ gives

\vspace{-0.1cm}

$$
\left(\frac{dV}{d\varphi}\right)_{\varphi = <\!\varphi\!>_{\rm vac}} \!\!\!\!\equiv <\!\!\varphi\!\!>_{\rm vac}\left(-\mu^2 + \lambda <\!\!\varphi\!\!>_{\rm vac}^2\right) = 0 \;\;,\;\; <\!\!\varphi\!\!>_{\rm vac} = \frac{\mu^2}{\lambda}
$$

\vspace{-0.1cm}

\ni (in the tree approximation). For $F_{\mu\,\nu}$ and $A_{\mu\,\nu}$, the Lagrangian $\cal{L}$ leads to the field equations

\vspace{-0.1cm}

$$
\partial^\nu(F_{\mu \nu} + \sqrt{f} \varphi A_{\mu \nu}) = - j_\mu \,,
$$

\vspace{-0.2cm}

\ni and

\vspace{-0.2cm}

$$
\left(\Box -  \frac{1}{2} \eta \,\varphi^2\right) A_{\mu \nu} = - \sqrt{f}\left(\varphi F_{\mu \nu} \!+\! \sum_i\zeta_i \bar{\psi}_i \sigma_{\mu\,\nu} \psi_i\right) \,,
$$

\ni with $F_{\mu \nu} = \partial_\mu A_\nu - \partial_\nu A_\mu$. They may be called {\it supplemented Maxwell equations}. Here, $F_{\mu \nu}$ and $j_\mu$ are the \SM electromagnetic field and electric current ($F_{\mu \nu}$ appears after the electroweak symmetry is spontaneously broken by the \SM Higgs bosons and photons emerge). The masses, {\it spontaneously generated} by $<\!\!\varphi\!\!>_{\rm vac}$, are $m_{\psi_i} \equiv \xi_i <\!\!\varphi\!\!>_{\rm vac}$, $m_\varphi \equiv \sqrt{2\lambda}<\!\!\varphi\!\!>_{\rm vac}$ and $M_A \equiv \sqrt{\eta/2} <\!\!\varphi\!\!>_{\rm vac}$. The sterino magnetic moments, also {\it spontaneously generated} by 
$<\!\!\varphi\!\!>_{\rm vac}$, are given through the field equation for $\psi_i$ following from the Lagrangian $\cal{L}$,

\vspace{-0.1cm}

$$
\left(i \gamma^\lambda \partial_\lambda \!-\! \xi_i \varphi \!-\! \frac{1}{2}\sqrt{f} \,\zeta_i \sigma_{\mu \nu} 
A^{\mu \nu} \right)\psi_i = 0 \,,
$$

\ni which in the recoil-free approximation of $A_{\mu \nu} \simeq [2\sqrt{f}/(\eta <\!\!\varphi\!\!>_{\rm vac})] 
F_{\mu \nu} = [2\sqrt{f} \xi_i/(\eta\,m_{\psi_i})]F_{\mu \nu}$ (with $\varphi \simeq <\!\!\varphi\!\!>_{\rm vac}$, $\psi_i \simeq 0$ and $\Box \simeq 0$) implies that

\vspace{-0.1cm}

$$
\frac{1}{2} \sqrt{f} \zeta_i \sigma_{\mu \nu} A^{\mu \nu} \simeq [{f}\,\zeta_i\,\xi_i/(\eta\,m_{\psi_i})] \sigma_{\mu \nu}F^{\mu \nu} \equiv \mu_{\psi_i}\sigma_{\mu \nu} F^{\mu \nu} \,.
$$

\ni Here, $\mu_{\psi_i} \equiv f\zeta_i \xi_i/(\eta\,m_{\psi_i})$ are the sterino magnetic moments. They are spontaneously generated for electrically neutral sterinos.}}

In such a model, the thermal freeze-out relic of sterinos in two or more generations is a candidate for cold dark matter. This requirement constrains strongly masses of sterinos, sterons and quanta of the field $A_{\mu\,\nu}$ ({\it cf.} Refs. [4] for the option with one sterino generation). In the simplest case, sterinos in two or more generations are stable, leading to stable cold dark matter in two or more species. But, some transitions between sterino generations, if allowed, may produce the exciting cold dark matter instead. 

The aim of the present note is to invoke a previous work [5] and show that the existence in Nature of {\it two or three generations} may be natural for some fundamental sterile spin-1/2 fermions, whereas  {\it three generations} are natural for \SM leptons and quarks. At any rate, such multiplicities of generations appear necessarily if the {\it fundamental matter particles} are formed in Nature in consistency with {\it Dirac's square root} procedure

\begin{equation}
\sqrt{p^2}\rightarrow \Gamma^\mu p_\mu\,,
\end{equation}

\ni where the Dirac algebra

\begin{equation}
\left\{ \Gamma^\mu\,,\,\Gamma^\nu \right\} = 2 g^{\mu \nu}
\end{equation}

\ni  is satisfied, and if  an {\it intrinsic Pauli principle} holds. The first requirement implies for free fundamental matter particles the wave equation in the general form of free Dirac equation

\begin{equation}
(\Gamma \cdot p - M) \psi(x) = 0 \,.
\end{equation}

\ni As we shall see, the second requirement will restrict the form of wave function $\psi(x)$.

\vspace{0.3cm}

\ni {\bf 2. Intrinsically composite model for fundamental matter particles}

\vspace{0.3cm}

To justify our claim, note that the general Dirac matrices $\Gamma^\mu$ satisfying Eq. (2) can be represented as $\Gamma^\mu \equiv\Gamma^{(N)\,\mu}$, where

\begin{equation}
\Gamma^{(N)\,\mu} \equiv \frac{1}{\sqrt{N}} \sum^N_{i=1} \gamma^{(N)\,\mu}_{i } \;\;\;(N = 1,2,3,\ldots)
\end{equation}

\ni are built up additively from the $4N\times 4N$ matrices $\gamma^{(N)\,\mu}_i$ being elements of the following Clifford algebra:

\begin{equation}
\left\{ \gamma^{(N)\,\mu}_i\,,\,\gamma^{(N)\,\nu}_j  \right\} = 2 g^{\mu \nu} \delta_{i j}\;\;\; (i,j = 1,2,\ldots,N)\,. 
\end{equation}

\ni Then, the free Dirac equation (3) becomes [5]

\begin{equation}
\left( \Gamma^{(N)}\cdot p -  M^{(N)}\right) \psi^{(N)}(x) = 0 \;\;\;(N = 1,2,3,\ldots)\,.
\end{equation}

\ni For $ N = 1$, Eq. (6) is the ordinary Dirac equation, while for $ N = 2$ it is known as the Dirac form [6] of  K\"{a}hler equation [7]. For $ N = 3,4,5,\ldots$ it gives us new Dirac-type equations [5]. When $N$ is odd or even, it describes spin-half-integer or spin-integer particles, respectively (for $ N = 2,3,4,\ldots$ it is spin-reducible).

Now, in place of the component Clifford matrices $\gamma^{(N)\,\mu}$ $(i = 1,2,\ldots, N)$ we can use their Jacobi-type combinations $\Gamma^{(N)\,\mu}_i\;\; (i = 1,2,\ldots, N)$ defined as

\begin{eqnarray}
\Gamma^{(N)\,\mu}_1 & \equiv & \frac{1}{\sqrt{N}} \left(\gamma^{(N) \mu}_{1} + \gamma^{(N) \mu}_{2} + \ldots + \gamma^{(N) \mu}_{N} \right) \equiv \Gamma^{(N)\,\mu} \,, \nonumber \\ 
\Gamma^{(N)\,\mu}_i & \equiv & \frac{1} {\sqrt{i(i - 1)}} \left[ \gamma^{(N) \mu}_{1} + \gamma^{(N) \mu}_{2} + \ldots + \gamma^{(N) \mu}_{i-1} - (i - 1) \gamma^{(N) \mu}_i \right] \nonumber \\ 
 & & \;\;\;\;\;(i = 2,3,\ldots, N) \,.
\end{eqnarray}

\ni They form the Clifford algebra

\begin{equation}
\left\{ \Gamma^{(N)\,\mu}_i\, ,  \,\Gamma^{(N) \nu}_j \right\} = 2g^{\mu \nu} \delta_{i j} \;\;\; (i,j = 1,2,\ldots,N)\;,
\end{equation}

\ni isomorphic with the Clifford algebra (5) of $\gamma^{(N)\,\mu}_i\;\;\; (i = 1,2,\ldots,N)$\,. 

Notice that all Jacobi-type spin and chiral matrices,

\begin{equation}
\Sigma^{(N) \mu \nu}_j  \equiv  \frac{i}{2} \left[ \Gamma^{(N) \mu}_j \, , \, \Gamma^{(N) \nu}_j  \right] \;\;\;\;,\;\;\;\; \Gamma^{(N)\,5}_j  \equiv  i \gamma^{(N) 0}_j \Gamma^{(N) 1}_j \Gamma^{(N) 2}_j  \Gamma^{(N) 3}_j  \,,
\end{equation}

\ni commute. They satisfy the relations

\begin{equation}
[\Sigma^{(N) k}_j \,,\, \Sigma^{(N) l}_j] = 2i \varepsilon^{k l m} \Sigma^{(N) m}_j \;\;\,,\;\;\, \left(\Gamma^{(N)\,5}_j\right)^2 =  {\bf 1}^{(4N,4N)} \,,    
\end{equation}

\ni where

\begin{equation} 
\Sigma^{(N) k l}_j \equiv \varepsilon^{k l m} \Sigma^{(N) m}_j \;\;\,,\;\;\,  \Sigma^{(N) 0l}_j  \equiv i \Gamma^{(N) 5}_j \Sigma^{(N) l}_j\;\; (k,l = 1,2,3) \,. 
\end{equation}

\ni Making use of the chiral representation of all Jacobi-type matrices $\Gamma^{(N) \mu}_i \;\;(i = 1,2,\ldots,N)$, where all $\Sigma^{(N) 3}_i$ and $\Gamma^{(N) 5}_i \;\;(i = 1,2,\ldots,N)$ are diagonal, we can reduce the general free Dirac equation (6) to the form

\begin{equation}
\left( \gamma \cdot p  - M^{(N)}\right)_{\alpha_1 \beta} \psi^{(N)}_{\beta \alpha_2 \ldots \alpha_N}(x) = 0
\end{equation}

\ni with $\gamma^\mu$ being the ordinary Dirac matrices, while 

\begin{equation}
\psi^{(N)}(x) = \left(\psi^{(N)}_{\alpha_1 \alpha_2 \ldots \alpha_N}(x)\right)\;.
\end{equation}

\ni In the wave function $\psi^{(N)}_{\alpha_1 \alpha_2 \ldots \alpha_N}(x)$ the numbers $\alpha_i = 1,2,3,4 \;(i = 1,2,\ldots,N)$ denote $N$ Dirac bispinor indices in the chiral representation of all $\Gamma^{(N) \mu}_i \;(i = 1,2,\ldots,N)$. We can see that $\alpha_1$ may be called the "centre-of-mass"\, bispinor index and $\alpha_2 \,,\, \ldots\,,\, \alpha_N$ --- the "relative"\, bispinor indices. Evidently, both notions have an intrinsic character, because the constituents of fundamental particles, appearing here, are the bispinor indices $\alpha_1 \,,\, \alpha_2 \,,\, \ldots \,,\, \alpha_N$ being algebraic objects rather than spatial ones ({\it cf.} Eq. (13)).

\vspace{0.4cm}

\ni {\bf 3. Fundamental matter particles with \SM gauge charges}

\vspace{0.4cm}

The \SM gauge interactions can be introduced into the general free Dirac equation (6) or (12) through the minimal substitution $p \rightarrow p - g A(x)$, where $p$ plays the role of the "centre-of-mass"\, four-momentum, while $x$ represents the "centre-of-mass"\, position. Then, the arising wave equation can be reduced  in our chiral representation to the form [5]

\begin{equation}
\left\{ \gamma \cdot \left[p - g A(x)\right] - M^{(N)}\right\}_{\alpha_1\beta} \psi^{(N)}_{\beta \alpha_2 \ldots \alpha_N}(x) = 0\;,
\end{equation}

\ni where $g \gamma \cdot A(x)$ symbolizes the \SM gauge coupling. It involves within $A(x)$ the familiar weak-isospin and color matrices, the weak-hypercharge dependence as well as the ordinary Dirac chiral matrix $\gamma^5 \equiv i \gamma^0 \gamma^1 \gamma^2 \gamma^3$  [8]. The \SM labels are suppressed in our notation.   

We can see that in Eq. (14) the "centre-of-mass" \,bispinor index $\alpha_1$ is distinguished from all "relative" \,bispinor indices $\alpha_2, \ldots, \alpha_N$ both by its dynamical coupling to the \SM gauge charges and its kinetic coupling to the four-momentum $p$. On the other hand, all $\alpha_2, \ldots, \alpha_N$ indices, being uncoupled, are mutually undistinguishable. Thus, it is natural to postulate that the "relative"\, bispinor indices  $\alpha_2, \ldots, \alpha_N$, treated as intrinsic physical objects, obey the Fermi statistics along with the {\it intrinsic Pauli principle} requiring {\it full antisymmetry} of matter wave function $\psi^{(N)}_{\alpha_1 \alpha_2 \ldots \alpha_N}(x)$ with respect to all $\alpha_2, \ldots, \alpha_N$ indices. 

Then, however, {\it only three} sorts of fundamental matter particles corresponding to the wave functions

\begin{equation}
\psi^{(1)}_{\alpha_1}(x)\;\;,\;\; \psi^{(3)}_{\alpha_1\alpha_2 \alpha_3}(x) \equiv \left(\gamma^5  C\right)_ {\alpha_2 \alpha_3} \psi^{(3)}_{\alpha_1}(x)  \;\;,\;\; \psi^{(5)}_{\alpha_1 \alpha_2 \alpha_3 \alpha_4 \alpha_5}(x) \equiv \varepsilon_{\alpha_2 \alpha_3 \alpha_4 \alpha_5} \psi^{(5)}_{\alpha_1}(x) 
\end{equation}

\ni can be realized for half-integer spin, and {\it only two}, corresponding to 

\begin{equation} 
\psi^{(2)}_{\alpha_1 \alpha_2}(x) \;\;,\;\; \psi^{(4)}_{\alpha_1 \alpha_2 \alpha_3 \alpha_4 }(x) \equiv \varepsilon_{\alpha_2 \alpha_3 \alpha_4 \alpha_5} \psi^{(4)}_{\alpha_1 \alpha_5}(x) \;,
\end{equation}

\ni for integer spin (here, $\left(\gamma^5 C\right)^T = - \gamma^5 C$). The former carry always spin 1/2,  the latter can have {\it a priori} spin 0,1,2. 

This result has been suggested [5] as a fundamental argument for the existence in Nature of {\it three and only three} lepton and quark generations. It is also an argument for there being in Nature {\it two and only two} generations of fundamental matter particles with integer spin and \SM charges like those of leptons and quarks. If the \SM Higgs boson is not a matter particle --- being rather a messenger particle like \SM gauge bosons --- it {\it does not belong} to either of these two matter generations with integer spin.

\vspace{0.4cm}

\ni {\bf 4. Sterile fundamental matter particles} 

\vspace{0.4cm}

When the "centre-of-mass" \,bispinor index $\alpha_1$ of $\psi^{(N)}_{\alpha_1 \alpha_2 \ldots \alpha_N }(x)$ is not dynamically coupled to the \SM gauge charges ({\it i.e.}, the corresponding sterile particles carry no such charges), then the matter wave equation takes the form (12), where the $\alpha_1$ index is distinguished from all $\alpha_2, \ldots, \alpha_N$ indices only by its kinematic coupling to the four-momentum $p$. Then, the previous requirement of full antisymmetry valid for all $\alpha_2, \ldots, \alpha_N$ can be extended also to $\alpha_1$, if and only if Eq. (12) is replaced by a form, where kinematic coupling is fully antisymmetrized with respect to all $\alpha_1, \alpha_2, \ldots, \alpha_N$, leading to the wave equation

\begin{equation}
 \stackrel{\rm{\Large antisym}}{~_{\alpha_1 \alpha_2 \ldots \alpha_N}}\left( \gamma \cdot p - M^{(N)}\right)_{\alpha_1\beta} \psi^{(N)}_{\beta \alpha_2 \ldots \alpha_N}(x) = 0\;.
\end{equation}

\vspace{0.2cm}

\ni In particular, for $N = 3$ Eq. (17) takes the form

\begin{equation}
\left(\gamma \cdot p \!-\! M^{(3)}\right)_{\!\!\alpha_1 \beta} \psi^{(3)}_{\beta \alpha_2 \alpha_3}(x) \!+\! \left( \gamma \cdot p \!-\! M^{(3)}\right)_{\!\!\alpha_2\beta} \psi^{(3)}_{\beta \alpha_3 \alpha_1}(x) \!+\! \left( \gamma \cdot p \!-\! M^{(3)}\right)_{\!\!\alpha_3\beta} \psi^{(3)}_{\beta \alpha_1 \alpha_2}(x)  = 0 \;, 
\end{equation}

\vspace{0.1cm}

\ni as $\psi^{(3)}_{\beta \alpha \alpha'}(x) = -\psi^{(3)}_{\beta \alpha' \alpha}(x)$ {\it ab initio}. Since $\alpha_1$ does not differ physically from all $\alpha_2, \ldots, \alpha_N$ in the absence of its dynamical coupling, it is natural to postulate for sterile matter particles the {\it full antisymmetry} of matter wave function $ \psi^{(N)}_{\alpha_1 \alpha_2 \ldots \alpha_N}(x)$ with respect to all $\alpha_1\,,\, \alpha_2 \,,\, \ldots \,,\, \alpha_N$ indices, and hence, our wave equation in the form (17).

In this case, {\it only two} sorts of fundamental matter particles corresponding to the wave functions

\begin{equation}
\psi^{(1)}_{\alpha_1}(x)\;\;,\;\; \psi^{(3)}_{\alpha_1\alpha_2 \alpha_3}(x) \equiv \varepsilon_{\alpha_1 \alpha_2 \alpha_3 \alpha_4} \psi^{(3)}_{\alpha_4}(x) 
\end{equation}

\vspace{0.1cm}

\ni can exist for half-integer spin, and also {\it only two}, corresponding to 

\begin{equation} 
\psi^{(2)}_{\alpha_1 \alpha_2}(x) \equiv (\gamma^5 C)_{\alpha_1 \alpha_2}\psi^{(2)}(x)  \;\;,\;\; \psi^{(4)}_{\alpha_1 \alpha_2 \alpha_3 \alpha_4 }(x) \equiv \varepsilon_{\alpha_1 \alpha_2 \alpha_3 \alpha_4} \psi^{(4)}(x) \;,
\end{equation}

\vspace{0.2cm}

\ni for integer spin. They carry always spin 1/2 and spin 0, respectively.

We have presented above a picture of sterile fundamental particles as it might look like, if these particles were nonactive not only with regard to the \SM gauge interactions, but also to all other possible interactions including familiar gravity and hypothetic hidden-sector interactions. Assuming that in our intrinsic model these extra interactions exist effectively and are coupled to the \,"centre-of-mass" \,\,degrees of freedom, we get the situation, where the \,"centre-of-mass" \,\,bispinor index $\alpha_1$ is always distinguished from all \,"relative"\,\, bispinor indices  $\alpha_2 \,,\, \ldots \,,\, \alpha_N$ and the latter are mutually undistinguishable. Then, there exist {\it three generations} of sterile spin-$1/2$ fermions and {\it two generations\,} of sterile spin-0,1,2 bosons, corresponding to the wave functions of the form (15) and (16), respectively. In this case, the wave functions (19) and (20) are not realized. We can believe that this interacting picture of sterile fundamental particles is more realistic. In a previous work [5], we assumed for them the noninteracting picture, where Eqs (19) and (20) hold.

\vfill\eject

\vspace{0.4cm}

\ni {\bf 5. Conclusions} 

\vspace{0.4cm}

The fundamental matter particles with spin $1\!/2$ and spin 0 described in Section 4 are sterile with regard to the \SM gauge charges and so, belong to the hidden sector of the Universe. Two or three former, fermions with spin 1/2,  if stable, may be responsible for the cold dark matter in two or three species, formed up as their thermal freeze-out relic. If the higher-mass species are not necessarily stable, they may lead to the exciting cold dark matter. Our sterinos in two or three generations, discussed in Introduction, {\it can be identified} with these two or three sterile matter particles with spin $1\!/2${\footnote {In one of earlier papers [9], we considered the option, where two light sterile neutrinos, possibly needed to explain the not null LSND effect, were identified with these sterile matter particles with spin $1\!/2$.}}. Two latter of the sterile matter particles, bosons with spin 0, if unstable (as may be expected), cannot provide the stuff for the cold dark matter. If our steron, also considered in Introduction, is not a matter particle --- being rather a messenger particle like, probably, the \SM Higgs boson --- it {\it cannot be identified} with either of these two sterile matter particles with spin 0.

However, we would like to emphasize that the results (15), (16) and (19), (20) --- which demonstrate, under our requirements, the existence of specific multiple generations of fundamental matter particles both with and without \SM gauge charges --- are obtained {\it independently} of our model of hidden sector [4] involving sterinos, sterons and quanta of the mediating field $A_{\mu \nu}$.

An essential ingredient of our model is the sterile mediating field $A_{\mu \nu}$  which interacts weakly with the electromagnetic field $F_{\mu \nu}$ through the bilinear form $\sqrt{f}(\varphi F_{\mu \nu} + \sum_i \zeta_i \bar{\psi}_i \sigma_{\mu \nu} \psi_i)$ (the source at $A_{\mu \nu}$), and so opens the photonic portal between \SM and hidden sectors. This portal is connected with sterino magnetic moments spontaneously generated by 
$<\!\!\varphi\!\!>_{\rm vac} \neq 0$ and coupled to $F_{\mu \nu}$. One may mention that the phrase "shedding light on dark matter", used neatly in the title of Ref. [10] in connection with the hypothetic magnetic moment of cold dark matter, characterizes {\it literatim} the situation in our model of hidden sector [4], where a narrow photonic portal allows for a weak transmission of light to and from the electrically neutral dark matter.

\vfill\eject

\vspace{0.3cm}

{\centerline{\bf References}}

\vspace{0.3cm}

{\everypar={\hangindent=0.65truecm}
\parindent=0pt\frenchspacing

~[1]~~G.B. Gelmini,  arXiv: 0810.3733 [{\tt hep-ph}]  ({\it 34th Int. Conf. on High Energy Physics}, Philadelphia, 2008).

~[2]~~N. Arkani-Hamed and N. Weiner,  arXiv: 0810.0713 [{\tt hep-ph}]; M.~Fairbairn and J.~Zupan,
 arXiv: 0810.4147 [{\tt hep-ph}]; and references therein. 
 
~[3]~~D.P. Finkbeiner and N. Weiner, {\it Phys. Rev.} {\bf D 76}, 083519 (2007); I.~Cholis, L.~Good\-enough and N.~Weiner, arXiv: 0802.2922 [{\tt astro-ph}].  

~[4]~~W. Kr\'{o}likowski, {\it Acta Phys. Polon.} {\bf B 39}, 1881 (2008) (arXiv: 0712.0505 [{\tt hep-ph}]); arXiv: 0803.2977 [{\tt hep-ph}];  arXiv: 0806.2036 [{\tt hep-ph}]; arXiv: 0809.1931 [{\tt hep-ph}].   

~[5]~~W. Kr\'{o}likowski, {\it Acta Phys. Polon.} {\bf B 21}, 871 (1990);  {\it Phys. Rev.} {\bf D 45}, 3222 (1992); {\it Acta Phys. Polon.} {\bf B 33}, 2559 (2002); {\tt hep--ph/0504256}; {\tt hep--ph/0604148}; {\it Acta Phys. Polon.} {\bf B 38}, 3133 (2007); and references therein. 
 
~[6]~~T. Banks, Y. Dothan and D.~~Horn, {\it Phys. Lett.} {\bf B 117}, 413 (1982).

~[7]~~E. K\"{a}hler, {\it Rendiconti di Matematica} {\bf 21}, 425 (1962); {\it cf.} also D.~Ivanenko and L.~Landau, {\it Z. Phys.} {\bf 48}, 341 (1928).

~[8]~~Particle Data Group, {\it Review of Particle Physics, Phys. Lett.} {\bf B 667}, 1 (2008).

~[9]~~W. Kr\'{o}likowski, {\it Acta Phys. Polon.}, {\bf B 31} 1913 (2000).

[10]~~S. Gardner, arXiv: 0811.0967 [{\tt hep-ph}].

\vfill\eject

\end{document}